\documentclass[11pt,prd,nofootinbib]{revtex4} 
\usepackage{graphicx} 
\usepackage{amsmath}
\usepackage{amsfonts,amsbsy}
\usepackage{amssymb}

\begin{document}

\begin{flushright}
LU TP 15-??\\
November 2015
\vskip1cm
\end{flushright}

\title{Tetraquark Production in Double Parton Scattering}
\author{F. Carvalho$^1$, E.R.  Cazaroto$^2$ , V.P. Gon\c{c}alves$^{3,4}$ and F.S. Navarra$^2$}

\affiliation{$^1$ Departamento de Ci\^encias Exatas e da Terra, Universidade Federal de S\~ao Paulo, 
Campus Diadema, Rua Prof. Artur Riedel, 275, Jd. Eldorado, 09972-270, Diadema, SP, Brazil \\
$^2$ Instituto de F\'{\i}sica, Universidade de S\~{a}o Paulo, CEP 05315-970 S\~{a}o Paulo, SP, Brazil\\
$^3$ Department of Astronomy and Theoretical Physics, Lund University, 223-62 Lund, Sweden.\\
$^4$ Instituto de F\'{\i}sica e Matem\'atica, Universidade Federal de Pelotas, CEP 96010-900, 
Pelotas, RS, Brazil.}

\begin{abstract}
We develop a model to  study  tetraquark production in hadronic collisions. We focus on double 
parton scattering and formulate a version of the color evaporation model for the production of the  
$X(3872)$ and of  the $T_{4c}$ tetraquark, a state composed by  the $c \bar{c} c \bar{c}$ quarks. 
We find that the production cross section grows rapidly with the collision energy $\sqrt{s}$ 
and make predictions for the forthcoming higher energy data of the LHC. 

\end{abstract}

\maketitle

\section{Introduction}

\vspace{0.5cm}

\subsection{Production mechanism}

Over the last years the existence of exotic hadrons has been firmly established \cite{espo,nnl} and now the 
next step is to determine their structure. Among the proposed configurations the  meson molecule 
and the tetraquark are the most often discussed. 
So far almost all the experimental information about these states comes from 
their production in B decays. The production of exotic particles in proton proton collisions is one 
of the most promising testing grounds for our ideas about the structure of the new states. It has 
been shown \cite{espo} that it is extremely difficult to produce molecules in p p collisions. In the 
molecular approach the estimated cross section for $X(3872)$ production is two orders of magnitude 
smaller than the measured one.  The present challenge for theorists is  to show that these data can 
be explained by the tetraquark model. To the best of our knowledge, this has not been done so far.  
In this work we give a  step in this direction, considering the production of the $X(3872)$ and  
of the $T_{4c}$, a state composed by two charm quark pairs: $c \bar{c} c \bar{c}$.

In recent high energy collisions at the LHC, it became relatively  easy to 
produce \cite{2psi7,2psilhcb}  four charm quarks ($c \bar{c} c \bar{c}$) in the same event. 
Events with four heavy quarks can be treated as a 
particular case of  $\alpha_s^2$ correction
to the standard single gluon-gluon scattering, in which an extra $c \bar{c}$ pair is produced, i.e., the process 
$ g g \rightarrow c \bar{c} c \bar{c} $. This is usually  called  single parton scattering (SPS). 
Another possible way to produce $c \bar{c} c \bar{c}$ is by two independent leading order gluon-gluon 
scatterings, i.e. two times the reaction $ g g \rightarrow c \bar{c}$. This is usually called double 
parton scattering (DPS) \cite{dps1,dps2}. In fact, apart from $c \bar{c} c \bar{c}$, DPS events may generate many other 
different final states, such as four jets, a $c - \bar{c}$ pair plus two jets, etc. For our purposes, the other 
relevant DPS process is the production of a $c - \bar{c}$ pair plus a light quark pair, $q - \bar{q}$, which will 
hadronize and form the $X(3872)$. Since DPS events are in the realm of perturbative physics, the light quark pair must 
be produced  with large invariant mass and the final state $X(3872)$ will carry large transverse momentum. This seems to 
be appropriate to describe the CMS data \cite{cms}, where the $X(3872)$ was observed with a transverse momentum lying in 
the range $10 \le p_T \le 25$ GeV.  In \cite{nosdps,anton} it has been shown that  DPS charm 
production is already comparable to SPS production at LHC energies.
DPS grows faster with the energy because it is proportional to $g(x,\mu^2)^4$  while SPS is 
proportional to $g(x,\mu^2)^2$. Here  $g(x,\mu^2)$ is the gluon density in the proton as a function of 
the gluon fractional momentum   $x$ and 
of the scale $\mu$ and it  grows quickly with increasing collision energies. 
In the present work we shall consider the DPS events  with the production of the 
two $c \bar{c}$ pairs and also with a $c \bar{c}$ and a light quark $q \bar{q}$ pair.  

\begin{figure}[t]

\begin{center}

\includegraphics[width=10.cm]{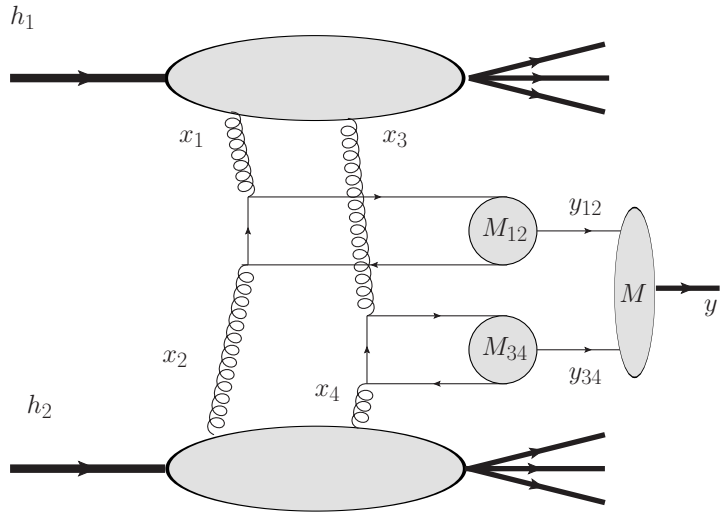}

\end{center}

\caption{The gluons with odd (even) label come from the upper (lower) hadron, and carry momentum 
fraction $x_i$.  The ``gluon 1'' scatters with  ``gluon 2'', making the state $M_{12}$. An analogous 
process occurs with gluons 3 and 4. Finally $M_{12}$ and $M_{34}$ merge and form the $T_{4c}$ with 
mass $M$.}

\label{fig1}

\end{figure}

Once we have generated all the quarks and antiquarks needed to form the $X(3872)$ or the $T_{4c}$ in DPS events, 
we need to bind them together. To this end we shall use  the main ideas of the Color Evaporation Model (CEM) \cite{cem,ramona}
of charmonium production, where the $c - \bar{c}$ is ``kinematically bound'', i.e., the charm pair sticks together 
because it does not have an invariant mass large enough to produce anything else.  
We shall use  the CEM ideas  to study $T_{4c}$ and $X(3872)$ production in DPS events.  
In the CEM formalism one parton from the hadron target scatters with one parton from the hadron 
projectile forming a charmonium state, which can absorb (emit) additional gluons from (to) the 
hadronic color field to become color neutral. This is the usual (SPS) $c \bar{c}$ production. At high energies the gluon density in 
the proton is much bigger than the sea quark density and hence, in what follows, we shall consider particle production only from 
gluon-gluon collisions. Now  we are going to extend the CEM to the case where two gluons from the hadron target scatter 
independently with two gluons from the hadron projectile as  depicted in  Fig. \ref{fig1}, where we 
show DPS production of $T_{4c}$.  In the figure two gluons collide and form a $c \bar{c}$ state with
mass $M_{12}$, while other two gluons collide and form a second $c \bar{c}$ state with mass $M_{34}$. 
The two objects bind to each other forming the $T_{4c}$. Additional gluon exchanges with the environment 
are not shown in  the figure. Replacing one $c \bar{c}$ pair by a light quark pair, $q \bar{q}$, the 
diagram would describe the production of $X(3872)$. 

The main difference between a tetraquark and  a meson molecule is that the former is compact and the interaction between the constituents occurs 
through color exchange forces whereas the latter is an extended object and the interaction between its constituents happens through meson exchange forces. 
In what was said above no explicit mention to size or color is made. However when we speak about the initial gluon fusion and about the final 
color neutralization through gluon emission or absorption it is understood that all this must happen within the confinement scale $\simeq 1$ fm. 
For this reason we believe that our model is suitable to describe tetraquark production. Although not explicitly excluded, it seems very unlikely 
that the clusters with masses $M_{12}$ and $M_{34}$ will form color singlets interacting through meson exchange.

\subsection{Kinematics}

Working with  the usual CEM  one-dimensional kinematics, the rapidities of the objects $M_{12}$ and 
$M_{34}$  are respectively:
\begin{eqnarray}
y_{12} \,\, = \,\, \frac{1}{2} \ln \frac{x_1}{x_2} \,\,\,\,\,\,\,\,  \mbox{and} \,\,\,\,\,\,\,\,  
y_{34} \,\, = \,\, \frac{1}{2} \ln \frac{x_3}{x_4}
\label{y12}
\end{eqnarray}
and their invariant masses are 
\begin{eqnarray}
M_{12} = \sqrt{x_1\,x_2\,s}   \,\,\,\,\,\,\,\,  \mbox{and} \,\,\,\,\,\,\,\,  
M_{34} = \sqrt{x_3\,x_4\,s} \,\,.
\label{m12}
\end{eqnarray}
In terms of these variables and in the low $p_T$ regime, the invariant mass of the 
$c \bar{c} c \bar{c}$ system is then  given by:
\begin{eqnarray}
M^2 \, = \, M_{12}^2 \, + \, M_{34}^2 \, + \, 2 M_{12} \, M_{34} \, \cosh (y_{12} - y_{34})\,\,.
\label{m2}
\end{eqnarray}
The $cosh$ function grows very rapidly with the argument and hence even a modest rapidity difference 
between the two clusters with $M_{12}$ and $M_{34}$ will significantly increase the value of $M$. 
We will then assume that both clusters move with equal rapidity, i.e. $y_{12} = y_{34}$, and become bound 
to  each other, forming a system with mass:
\begin{equation}
M \, = \, M_{12} \, + \, M_{34}\,\,.
\label{msum}
\end{equation}
Finally, in order to produce the final tetraquark state with right mass, $M_T$, the cluster with mass  $M$ 
emits or absorbs gluons carrying an energy $\Delta$, which will be discussed below. We have thus:
\begin{equation}
M \pm \Delta \, = \, M_T \,\,.
\label{msumt}
\end{equation}
A remarkable difference between the standard CEM for charmonium production and the model developed here is in 
the role played by the limits of the integral over the squared invariant mass $M^2$. In the case of the usual $J/\psi$ 
production it goes from $(2m_c)^2$ to $(2m_D)^2$.  This ensures that the $c -\bar{c}$ can never decay into open charm, 
not forming the charmonium state, because it does not have enough invariant mass. The case of the tetraquark $X(3872)$ 
is different. Suppose, for example, that we have the four-quark system with an invariant mass of $3740$ MeV. While this system 
can only form the $X$ resonance by absorbing some gluons (carrying energy $\Delta$) from the target or from the projectile, 
it has sufficient mass to decay immediately into a $D^+ D^-$ pair and not form the resonance. Moreover, since the energy $\Delta$ is
carried by an undefined  number of gluons, this decay is not hindered by  parity (or charge conjugation) conservation.   
Therefore, in our case, the integration over $M^2$ must be changed becoming more restrictive:
\begin{equation}
\int_{(2m_c)^2}^{(2m_D)^2} \, d M^2   \,\, \rightarrow \,\, \int_{(M_T - \Delta)^2}^{(M_T + \Delta)^2} \, d M^2 
\label{redef}
\end{equation}
where the left side refers to the usual CEM and the right side refers to tetraquark states. We will use this restriction in Sec. III.

\section{Tetraquark  production}

\subsection{ $T_{4c}$: the all-charm tetraquark}

The $T_{4c}$  state was first discussed long time ago by Iwasaky \cite{iwasaky}. In the eighties and early nineties, many authors 
\cite{ader1,ht,brac,sema} addressed the subject arriving at different conclusions concerning the existence of a $c \bar{c} c \bar{c}$ 
bound state. More recently, with the revival of charmonium spectroscopy,
Lloyd and Vary \cite{vary} investigated the four-body $c \bar{c} c \bar{c}$ system  obtaining several close-lying
bound states. They found that deeply bound ($\simeq 100$ MeV)  states may exist with masses
around $6$ GeV.   In Ref. \cite{javier}  the existence of $c \bar{c} c \bar{c}$ states was discussed 
in the framework of the hyperspherical harmonic formalism. The results suggested the possible existence
of three four-quark bound states with quantum numbers $0^{+-}$, $2^{+-}$ and $2^{++}$ and masses
of the order of $6.50$, $6.65$, and $6.22$ GeV. More recently, using the Bethe-Salpeter approach, the
authors of Ref. \cite{heupel} found an all-charm tetraquark  with $J^{PC} =0^{++}$ and
mass $M_{T_{4c}} = 5.3 \pm 0.5$ GeV.
This mass is considerably lower than the $6.0$  GeV obtained in the previous model calculations
\cite{iwasaky,vary}. It is also much lower than the $2 \eta_c$ threshold. Potential decay channels into
D mesons and pairs of light mesons necessarily involve internal gluon lines. The resulting
decay width may therefore be rather small. On the other hand, preliminary lattice QCD calculations
\cite{wagner,bicudo} seem to disfavor the existence of a deeply bound  $c \bar{c} c \bar{c}$ state,
being more compatible with a loosely bound $2 \eta_c$ molecular state.
In the works \cite{russo1,russo2}  $T_{4c}$ production was studied in SPS events.

\subsection{The production cross section}


The cross section of the process shown in Fig.~\ref{fig1} can be calculated with  the 
schematic DPS ``pocket'' formula:
\begin{equation}
\sigma_{DPS} \propto  \frac{\sigma_{SPS}^{12} \sigma_{SPS}^{34}}{\sigma_{eff}}
\label{dpspock}
\end{equation}
where $\sigma_{eff} \simeq 15$ mb is a constant extracted from data analysis and $\sigma_{SPS}$ 
is the standard QCD parton model formula, i. e.,  the convolution of parton densities with partonic 
cross sections.  To be more precise we expand the above formula showing the kinematical constraints
introduced to study tetraquark production. It reads:
\begin{eqnarray}
\sigma_{DPS} &=& \frac{F_{T_{4c}}}{\sigma_{eff}}  
\left[  \int _0^1 dx_1 \int _0^1 dx_2 \,\,
g(x_1,\mu^2) \, g(x_2,\mu^2) \, \sigma _{g_1 g_2 \to c\bar{c}} \right] \nonumber \\
  & \times & \,\, 
\left[  \int _0^1 dx_3 \int _0^1 dx_4 \,\,
g(x_3,\mu^2) \, g(x_4,\mu^2) \, \sigma _{g_3 g_4 \to c\bar{c}} \right] \nonumber \\
  & \times & \,\, \Theta (1-x_1-x_3) \,\,\Theta (1-x_2-x_4) \,\, 
\Theta (M_{12}^2 - 4m_c^2) \,\, \Theta (M_{34}^2 - 4m_c^2) \nonumber \\
& \times & \,\, \delta(y_{34} - y_{12})
\label{sigtot}
\end{eqnarray}
where $g(x,\mu^2)$ is the gluon distribution in the proton with the gluon fractional momentum 
$x$ and at the factorization scale $\mu^2$ and $\sigma _{g g \to c\bar{c}}$ is the 
$g g\to c\bar{c}$ elementary cross-section. The  step functions $\Theta (1- x_1 - x_3)$ and
$\Theta (1- x_2 - x_4)$    enforce momentum conservation in the projectile and in the target. 
The step functions $\Theta (M_{12}^2 - 4m_c^2)$ and $\Theta (M_{34}^2 - 4m_c^2)$   guarantee 
that the invariant masses of the gluon pairs 12 and 34 are large enough to produce two charm 
quark pairs. The delta function implements the ``binding condition''  and $F_{T_{4c}}$ is a constant, 
analogous to the one appearing in the CEM formula, which represents the probability of the four-quark 
system to evolve to a particular tetraquark state. 

In the above formula, all the variables depend on the momentum fractions $x_1$ ... $x_4$. Because of 
the delta function, we know that the two clusters shown in Fig.~\ref{fig1} are ``flying together'' 
and that they form a system with mass $M = M_{12} + M_{34}$, which can take any value. In order to 
improve our kinematical description of this bound state, we can impose constraints on the values of 
$M$, such as (\ref{redef}). This can be best done rewritting (\ref{sigtot}) and changing 
variables from $x_1$, $x_2$, $x_3$ and $x_4$ to $y_{12}$, $y_{34}$, $M_{12}$ and $M_{34}$. We obtain:
\begin{eqnarray}
\sigma_{DPS} &=& \frac{F_{T_{4c}}}{\sigma_{eff}}
\left[ \frac{1}{s} \int  d y_{12} \int dM_{12}^2 \,\,
g(\bar{x_1},\mu^2) \, g(\bar{x_2},\mu^2) \, \sigma _{g_1  g_2 \to c\bar{c}} \right] \nonumber \\
  & \times & \,\,
\left[ \frac{1}{s} \int  dy_{34} \int  dM_{34}^2 \,\,
g(\bar{x_3},\mu^2) \, g(\bar{x_4},\mu^2) \, \sigma _{g_3 g_4 \to c\bar{c}} \right] \nonumber \\
  & \times & \,\, \Theta (1- \bar{x_1} - \bar{x_3}) \,\,\Theta (1- \bar{x_2} - \bar{x_4}) \,\,
\Theta (M_{12}^2 - 4m_c^2) \,\, \Theta (M_{34}^2 - 4m_c^2) \nonumber \\
& \times & \,\, \delta(y_{34} - y_{12})
\label{sigtotym}
\end{eqnarray}
where
\begin{equation}
\bar{x_1} = \frac{M_{12}}{\sqrt{s}} \, e^{y_{12}}   \,\, \hspace{0.3cm} , \hspace{0.3cm}
\bar{x_2} = \frac{M_{12}}{\sqrt{s}} \, e^{- y_{12}} \,\, \hspace{0.3cm} , \hspace{0.3cm}
\bar{x_3} = \frac{M_{34}}{\sqrt{s}} \, e^{y_{34}}   \,\, \hspace{0.3cm} , \hspace{0.3cm}
\bar{x_4} = \frac{M_{34}}{\sqrt{s}} \, e^{- y_{34}}
\label{redefex}
\end{equation}
and consequently
\begin{equation}
\Theta (1- \bar{x_1} - \bar{x_3}) = \Theta (1- \frac{M_{12}}{\sqrt{s}} \, e^{y_{12}}  
- \frac{M_{34}}{\sqrt{s}} \, e^{y_{34}} )  \,\,\hspace{0.3cm} , \hspace{0.3cm} \,\,
\Theta (1- \bar{x_2} - \bar{x_4}) = \Theta (1- \frac{M_{12}}{\sqrt{s}} \, e^{- y_{12}}
- \frac{M_{34}}{\sqrt{s}} \, e^{ - y_{34}} )
\label{tetared}
\end{equation}
From the above expressions it is easy to see that when $y_{12} = y_{34} = y$, then (\ref{msum}) holds 
and the theta functions give lower and upper limits for the integration in $y$:
\begin{equation}
- \mbox{ln} \frac{\sqrt{s}}{M} \le y \le  \mbox{ln} \frac{\sqrt{s}}{M}
\label{ylim}
\end{equation}
The upper limit of $M_{12}$ and $M_{34}$  can be fixed imposing constraints on their sum, $M$. 
In the case of the $X(3872)$ we already know the mass of the state that we want to produce. In 
principle we could just use (\ref{msum}) with a fixed value of $M$. However, following the spirit of 
the CEM, we will assume that when the system with mass $M = M_{12} + M_{34}$ goes to the final state with mass 
$M_T$ it can absorb or emit  soft gluons to neutralize color. These gluons carry  an energy going from
almost zero to the QCD scale, given by $ \Delta =  \mathcal{O} (\Lambda_{QCD})$. 
Then, from (\ref{msum}) and (\ref{msumt}) 
we have:
\begin{equation}
M^{min} = M^{min}_{12} + M^{min}_{34} = M_T - \Delta
\label{minf}
\end{equation}
and
\begin{equation}
M^{max} = M^{max}_{12} + M^{max}_{34} =  M_T + \Delta
\label{msup}
\end{equation}
From these equations we can see that, knowing the mass of the tetraquark state and fixing the amount of energy which can be
exchanged in the formation of the final state, we constrain the limits in the integrations over $M_{12}$ and $M_{34}$. In the 
symmetric case of $T_{4c}$ production $M_{12}^{min} = M_{34}^{min}$, $M_{12}^{max} = M_{34}^{max}$, (\ref{minf}) and (\ref{msup}) 
completely fix these limits. In the case of the $X(3872)$, we may have different choices for $M_{12}^{min} ( M_{12}^{max})$  and 
$ M_{34}^{min} (M_{34}^{max})$ but they will be correlated.

\section{Numerical results and discussion}

\vspace{0.5cm}

\subsection{$T_{4c}$}

As mentioned in the introduction we take the production cross section of the $T_{4c}$ as a baseline because it is heavy, and 
hence treatable in pQCD, and also to make some contact with the production of $c \bar{c} c \bar{c}$ in DPS.  In this subsection we discuss 
the numerical results obtained for $T_{4c}$. Then in the following subsection, after only a few changes  we calculate the cross section for $X(3872)$ 
production.

We now evaluate equation (\ref{sigtotym}) replacing $g(x,\mu^2)$ by the MRST gluon distribution 
\cite{mrst} and  $\sigma_{g g \rightarrow c \bar{c}}$ by the standard leading order QCD result \cite{ramona}:
\begin{equation}
\sigma_{g g \rightarrow c \bar{c}} = 
\frac{\pi \alpha_s^2(m^2)}{3 m^2} \left\{ {(1+\frac{4m^2_c}{m^2}+\frac{m^4_c}{m^4}) \, 
ln [\frac{1+\beta}{1-\beta}] \, - \, \frac{1}{4} (7 + \frac{31 m^2_c}{m^2}) 
\beta } \right\}
\label{sigramona}
\end{equation}
with
$$
\beta = [1-\frac{4 m^2_c}{m^2}]^{1/2}
$$
where $m^2$ is equal to $M^2_{12}$ or $M^2_{34}$.  A difficulty in our calculation is the uncertainty 
in the normalization  of the cross section. Whereas in the case of charmonium production in the CEM  
we have experimental information, which can be used to fix the nonperturbative constant 
$F_H$, in the case of the $T_{4c}$ nothing is known. For the time being we can only try to make a 
simple estimate.

In the usual CEM it is  assumed that the nonperturbative probability for the $Q \bar{Q}$ pair 
to evolve into a quarkonium state H is given by a constant $F_{H}$ that is energy-momentum and process 
independent. Once $F_{H}$ has been fixed by comparison with the measured total cross section for the 
production of the quarkonium H at one given energy, the CEM can predict, with no additional
free parameters, the energy dependence of the production cross section and the momentum distribution of 
the produced quarkonium. Following the CEM strategy we shall adjust $\sigma_{T_{4c}}$ 
connecting it to the experimentally  measured cross section of $X(3872)$ production at one single 
energy and then make predictions for higher energies. 

\begin{figure}[h]

\begin{center}
\vspace{0.1cm}
\includegraphics[width=9.0cm]{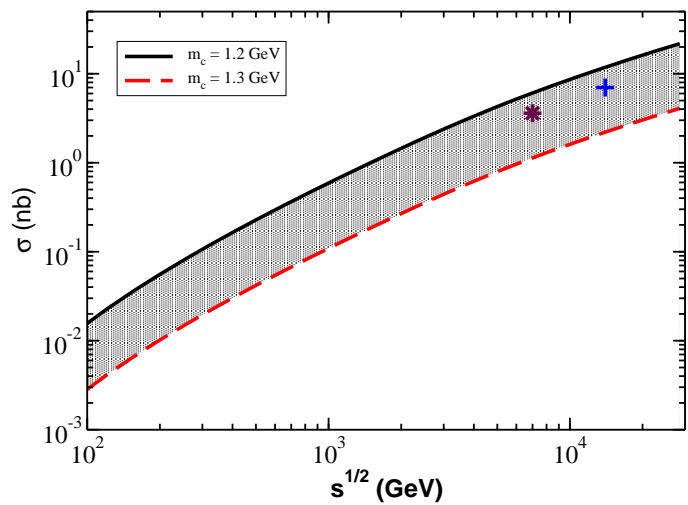}

\end{center}
\vspace{-0.2cm}
\caption{Cross section of $T_{4c}$ production as a function of the energy. }

\label{fig2}

\end{figure}

We know that the production cross section of $T_{4c}$  must be smaller than the one for $X(3872)$ 
production and the latter has been measured by the CMS collaboration \cite{cms} at $\sqrt{s} = 7$ TeV. Moreover, 
assuming that the binding mechanism is the same, the only difference is that we must replace the light 
quark pair (which is in the $X(3872)$) by the $c \bar{c}$ pair, which is much more difficult to produce.  
Therefore, in order to estimate the cross section for producing the $T_{4c}$, we must  multiply the 
$X(3872)$ production cross section, $\sigma_{X}$, by a penalty factor:
\begin{equation}
\sigma_{T_{4c}} = \frac{\sigma_{c \bar{c} c \bar{c}} }{\sigma_{c \bar{c} q \bar{q}}} \, \sigma_{X} \simeq 
\frac{\sigma_{c \bar{c}} \,  \sigma_{c \bar{c}} }{\sigma_{c \bar{c}} \, \sigma_{q\bar{q}}} \, \sigma_{X} 
\simeq \frac{\sigma_{c \bar{c}}}{\sigma_{inel}} 
\simeq 0.12 \, \sigma_{X}
\label{sigt4c}
\end{equation}
where $\sigma_{c \bar{c} c \bar{c}}$ and $\sigma_{c \bar{c} q \bar{q}}$ are the cross sections for 
the production of $c \bar{c} c \bar{c}$ and $c \bar{c} q \bar{q}$ respectively. These cross sections 
can be measured in double parton scattering events. In the above expression, after using the factorization 
hypothesis, $\sigma_{c \bar{c}}$ cancels out and the ratio  $\sigma_{c \bar{c}}/ \sigma_{q \bar{q}}  \simeq   \sigma_{c \bar{c}}/ \sigma_{inel}$ 
can be  inferred from data \cite{siginel,cc7}, which at $7$ TeV  yield $\approx 0.12$. All the required numbers are collected in Table I. 
Finally, using the value of $\sigma_{X} \simeq 30$ nb \cite{cms}, we have:
\begin{equation}
\sigma_{T_{4c}}(\sqrt{s} =  7 \, \mbox{TeV}) \simeq (3.6 \, \pm 2.5)  \, \mbox{nb}
\label{sigtf}
\end{equation} 
Having fixed the numbers we plot the cross section for $T_{4c}$ production as a function of the energy in Fig. \ref{fig2}. In order to obtain an 
estimate of the theoretical error we vary the parameters trying to scan the most relevant region in the parameter space. 
We choose $\Delta \approx \Lambda_{QCD} \approx 200$ MeV and we assume that the $T_{4c}$ mass is given by $M_{T_{4c}} = 5.4$ GeV, as obtained 
in Ref. \cite{heupel}. With these two parameters fixed we can choose different values for the charm mass $m_c$. 
However there is an upper limit for $m_c$, which cannot be bigger than $M_{12}^{min}/2$. Substituting $\Delta$ and $M_{T_{4c}}$ in 
Eq. (\ref{minf}) we find that $M_{12}^{min} = 2.6$ GeV, consequently the maximum value for $m_c$ is $m_c = 1.3$ GeV. In Fig. \ref{fig2} the upper 
line corresponds to $m_c=1.2$ GeV and the lower line corresponds to  
$m_c=1.3$ GeV. The star in  Fig. \ref{fig2} corresponds to the central value at $\sqrt{s} = 7$ TeV. Here the constant $F_{T_{4c}}$ was chosen so as to reproduce   
(\ref{sigtf}). Once all the parameters are fixed at $\sqrt{s} = 7$ TeV, the energy dependence of the cross section is completely determined by the model. 
In Fig. \ref{fig2} the cross represents the central value of our prediction for the energy  $\sqrt{s} = 14$ TeV:
\begin{equation}
\sigma_{T_{4c}}(\sqrt{s} =  14 \, \mbox{TeV}) \simeq (7.0 \, \pm 4.8) \,  \mbox{nb}
\label{predt4c}
\end{equation}
The main feature of the curves is the rapid rise with $\sqrt{s}$, which might render the $T_{4c}$ 
observable already at 14 TeV. This same fast growing trend was observed in other estimates with DPS 
\cite{dps1,dps2}. 
\begin{center}
\begin{table}[t]
\caption{$X(3872)$ and $T_{4c}$  production cross sections at $\sqrt{s}=7$ TeV and $\sqrt{s}=14$ TeV.}
\vspace{0.5cm}
\begin{center}
\begin{tabular}{| c c c c c|}
\hline
\hline
Energy (TeV)                   &  $\sigma_{c \bar{c}}$ (mb)  & $\sigma_{inel}$ (mb)   & $\sigma_X$ (nb)          & $\sigma_{T_{4c}}$ (nb)        \\
\hline 
7                              &  8.5  \cite{cc7}            &  73.2  \cite{siginel}  & 30.0 \cite{cms}          & $3.6  \, \pm \, 2.5$          \\
\hline
14                             &                             &                        &  $44.6 \, \pm \, 17.7$  \,\,   & $ 7.0 \, \pm \, 4.8$           \\
\hline 
\hline
\end{tabular}
\end{center}
\label{tab1}
\end{table}
\end{center}

\subsection{$X(3872)$}

\begin{figure}[h]
\begin{center}
\vspace{1.0cm}
\includegraphics[width=9.0cm]{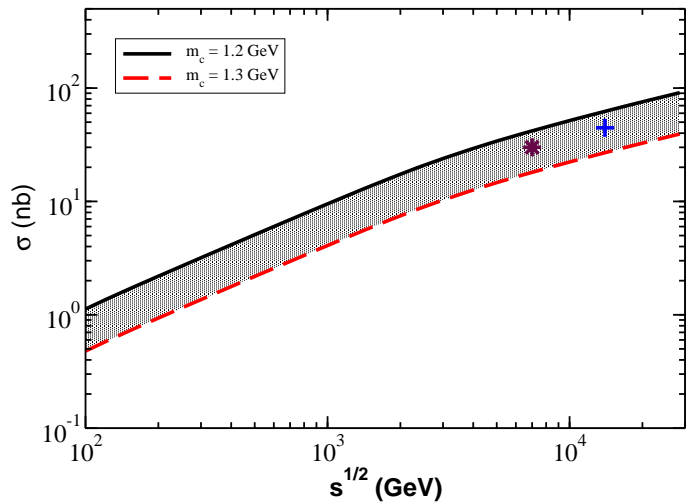}
\end{center}
\vspace{-0.1cm}
\caption{ $X(3872)$ production cross section as a function of the energy.}
\label{fig3}
\end{figure}

We now turn to the production cross section of $X(3872)$. We use the same parton densities as in the previous subsection and 
also the elementary cross section  for heavy quark production (\ref{sigramona}). Note that we use this expression even for 
light quark production $\sigma _{g_3 g_4 \to q \bar{q}}$, which appears now in the second line of  (\ref{sigtot}) or (\ref{sigtotym}). 
Since this expression only holds for heavy enough quarks, its use here is questionable. In spite of this uncertainty, the existing 
experience in the literature is encouraging. In \cite{guti} the authors used (\ref{sigramona}) to compute the cross section of strange particle 
production and calculated the asymmetries in the production of $K^+ / K^-$, $\Lambda/ \bar{\Lambda}$, ...etc. They have used the convolution formula  
of the parton model and have taken the strange quark mass to be $m_s \simeq 500$ MeV. They could reproduce well the existing data on asymmetries. In our 
case the invariant mass $M_{34}$ defines the perturbative QCD scale and hence we must have $M_{34} > 1$ GeV. This can be achieved with the light quarks having 
masses close to zero and transverse momenta  in the few GeV region. Since we are using the one-dimensional version of the CEM, instead of transverse momentum 
we will assign an effective mass to the light quarks, $m_q = 0.5$ GeV, which garantees that $M_{34} > 2 m_q > 1$ GeV. Moreover, choosing  $N_f=2$ and 
$\Lambda_{QCD}=200$ MeV,  we have typically:
\begin{equation}
\alpha_s =  \frac{12 \pi}{ (33 - 2 N_f) \,  \mbox{ln} ( \frac{(2 m_q)^2}{\Lambda_{QCD}^2})} \simeq 0.4
\label{alfs}
\end{equation}
Although we may expect significant corrections,  this number is still small enough for perturbation theory to make sense. 
As in the previous subsection, after fixing these parameters and knowing the tetraquark mass $M_X = 3872$ MeV the only remaining free parameters are the charm 
mass and the constant $F_X$. We show our results in Fig.~\ref{fig3}, where the upper line corresponds to $m_c=1.2$ GeV and the lower line corresponds to  
$m_c=1.3$ GeV. The constant $F_X$ was adjusted so that the central value of the cross section at $\sqrt{s} = 7$ TeV (shown with a star) corresponds to  
$\sigma_X = 30.0 $ nb. With all the numbers fixed at the lower energy the energy dependence is given by the model. At $\sqrt{s} = 14$ TeV, the cross indicates 
the central value of our prediction: 
\begin{equation}
\sigma_{X} (\sqrt{s} =  14 \, \mbox{TeV}) \simeq 44.6 \, \pm 17.7 \,  \mbox{nb}
\label{predx}
\end{equation}
The error in the number given above is relatively large but, at least we can predicit the order of magnitude of the cross section. As a first estimate with DPS, we 
think that the result is satisfactory.  The model presented here can be improved in several aspects. Probably the most relevant one is the prescription to form the 
resonance, i.e., the hadronization of the multiquark system. Progress in this direction would also benefit the SPS calculations of this process. Our prescription, 
based mostly on the kinematical aspects and using only the rapidities and invariant masses, is not accurate enough and is the largest source of uncertainties. Work 
along this line is in progress. The other sources of uncertainties are, as usual, the choice of parton densities, the choice of the energy scale at which they are 
computed, the choice of the scale at which $\alpha_s$ is computed, the choice of $\Lambda_{QCD}$, and the charm and light quark masses. 

\section{Conclusion} 

We have developed a model for tetraquark production which combines double parton scattering and  
the basic ideas of the color evaporation model. We have made predictions for the $X(3872)$ production 
cross section, which may be confronted with the forthcoming LHC data taken at $\sqrt{s} = 14$ TeV. 

The results presented above contain some uncertainties: i) they do not include tetraquark production in 
SPS  events, which can be larger than the DPS cross section. The calculation of the SPS cross section requires 
some fragmentation function which is not known.; ii) the binding mechanism is probably too simple and 
insensitive to the quantum numbers of the involved particles; iii) in the case of $X(3872)$ production, the use 
of formula (\ref{sigramona}) for light quark production is questionable. This problem may be circunvented
using the next-to-leading order version of the CEM, in which the transverse momentum is included. In this case
the light quarks can be really light but they have large $p_T$, rendering plausible the use of the perturbative 
formula (\ref{sigramona}).

\section*{Acknowledgements} 

We are grateful to J.M. Richard and to M. Nielsen for instructive discussions. 
This work was partially supported by the 
Brazilian funding  agencies CAPES, CNPq, FAPESP and FAPERGS.

\end{document}